# Nonlinear Doppler – Free comb-spectroscopy in counter-propagating fields

**S.A. Pulkin, A. Kalinichev, V.Arnautov, S.V. Uvarova, S. Savel'eva**
Department of General Physics of Physical Faculty, Saint-Petersburg State University, Saint-Petersburg, Russia
Email: spulkin@mail.ru

**Abstract**

The method of Doppler – free comb – spectroscopy for dipole transitions was proposed. The calculations for susceptibility spectrum for moving two-level atoms driving by strong counter propagating combs have been done. The used theoretical method based on the Fourier expansion of the components of density matrix on two rows on kv (v-velocity of group of atoms, k-projection of wave vector) and Ω (frequency between comb components). For testing of validity of this method the direct numerical integration was done. The narrow peaks with homogeneous width arise on the background of Doppler counter. The contrast of these peaks is large for largest amplitudes of comb-components. Power broadening is increasing with increase of field amplitudes. The spectral range of absorption spectrum is determined by the spectral range of comb generator and all homogeneous lines arise simultaneously. The spectral resolution is determined by the width of homogeneously –broadening lines. The physical nature of narrow peaks is in the existence of multi-photon transitions between manifolds of quasi-energy levels arising for different groups of atoms moving with velocities that satisfy to the resonant conditions   2kv=  (n+l)Ω, where n, l-are integers and Ω - frequency difference between comb teeth.

**Keywords:** Doppler – Free Comb Nonlinear Spectroscopy in Counter – Propagating Comb Laser Fields

## 1. Introduction

The goal of present work is the   research of susceptibility spectra in the strong counter-propagating polyharmonic light fields with phasing equidistant components. The frequency spectrum of this field is a spectrum of comb – generator of femtosecond laser. Generator radiates a time subsequence (train) of femtosecond pulses with repetition period of $\tau_p$. The radiation spectrum consists of narrow equidistant peaks (comb – spectrum). Nowadays comb – spectroscopy is a fast developing area of spectroscopy [1,2] allowing to detect with high sensitivity atomic and molecular lines in the wide spectral range with resolution limited by Doppler broadening for one –photon transitions. The main advantage of comb – spectroscopy method is in the possibility to detect simultaneously all spectral lines. The method of two – photon Doppler – free comb – spectroscopy has been developed in [3,4]. The all advantages of usual two- photon spectroscopy with advantages of comb – spectroscopy method are used in the given method, but the spectral resolution is determined by homogeneous broadening. In the present work we study the mechanism of arising of narrow peaks in the absorption spectrum at the background of Doppler shape of dipole transition in one photon excitation.

There is the interaction of counter – propagating waves with medium of moving atoms. Each molecule moving with velocity v "sees" not one field, but two fields, which detuned one from other on the frequency 2 kv. The first field with frequency $\omega_1$ always is in resonance with transition frequency, then on the frequency $\omega_{21} + 2$ kv in the frame of moving atom coordinate the second field acts. If both fields are strong the system converts from two-level to multi – level one consisting of manifold quasi –energy levels. The transitions between quasienergy levels are possible [6]. Two manifolds shifted on frequency 2kv arise for the case of moving atoms with velocity v. The narrow coherent peaks in the polarization and susceptibility spectrum    arise. This work is a prolongation of our previous work [7], where we considered the interaction of two counter propagating waves with two – level system – bi-harmonic (strong modulated) and weak counter – propagated monochromatic field. It was shown that in the polarization spectrum the additional narrow resonances arise on the frequencies that satisfy of the conditions of multi – photon resonances. We can explain the result of present researches using results of papers [6, 7]. The possibility of Doppler shift compensation for the first time proposed in our previous work [11], when direct integration of density matrix equations was made. There we showed that the mechanism of Doppler shift compensation in two-photon comb – spectroscopy and method described here are different. In the case of two-photon comb – spectroscopy one photon comes from low frequency side of comb – spectrum with frequency $\omega_0 - \Omega$ and second photon with frequency $\omega_0 + \Omega$ comes from high frequency side. The effective two – photon transition on transition frequency $\omega_{21} = 2\omega_0$ (where $\omega_0$ – carrier laser frequency, it coincides with the center of Doppler counter) will arise for the group of atoms when the resonant condition is satisfied: $kv = \Omega$. ($\Omega$ - is the frequency interval between comb – components). The other cause is in the case of counter – propagated combs acting near one – photon transitions [11]. The effective interaction for multi – photon transitions between quasi – energy levels on the frequency of one – photon transition takes place for groups of atoms with velocity v, This group of atoms interacts effectively with comb components on the frequencies $\omega_0 - p\Omega$ and $\omega_0+n\Omega$. (n and p –are integers and zero) for n = - p and when $2kv = n\Omega$.

Thus, as it follows from [11], the physical nature of narrow peaks is in the existence of multi-photon transitions between manifolds of quasi-energy levels arising for different groups of atoms moving with velocities that satisfy to the resonant conditions $2kv = (n+l)\Omega$, where n, l-are integers and $\Omega$ - frequency difference between comb teeth.

There is partial case when the strong modulated field and counter – propagating weak probe field act on Doppler broadened medium. This case is realized in the method of modulated transfer spectroscopy (MTS) [8, 9]. The four-wave mixing in medium leads to arising of modulation of probe field on exit of cell with atoms. The order of nonlinearity is determined in this case by nonlinear susceptibility $\chi(3)$. The condition of phase synchronism is satisfied in this case. The condition of phase control and synchronism is necessary to satisfy for multi-photon interactions in our case too.

## 2. Theory and Results of Calculations

### 2.1. Criteria for strong field

The probability of coherent nonlinear processes is determined by the amplitude of acting laser field.

For pulse excitation for arising of multi – photon processes is necessary that pulse square should be of order of π. Let's estimate the intensity (density of power) for pulse duration $\tau_p = 100$ fs and for energy $\xi = 10$ nJ for diameter of laser beam d = 1 mm:

$$I = 4\pi \frac{\xi}{\tau_p d^2} \approx 10^{11}(W/m^2)$$

$$Q = \int_0^t V(t)hdt \approx \Omega_R \tau_p \sim 1$$

So, for these parameters of laser pulse the laser field is strong and multi – photon transitions in the structure of quasi –energy levels are possible.

## Theory and results

The two row Fourier expansion method for density matrix elements [12] for the case of moving atoms was appied for solving density matrix equations.

The field member has the form:

$$E(t) = \frac{1}{2}\left\{W_{1m}e^{i(\omega_{s0}-kv)t} + W_{2m}e^{i(\omega_{p0}+kv)t} + c.c.\right\}$$

where $W_{1m}$, $W_{2m}$ – the comb-amplitudes of counter-propagating waves:

$$W_{1m} = \sum_{m=-n}^{n} E_{1m}e^{im\Omega t}, \quad W_{2m} = \sum_{m=-n}^{n} E_{2m}e^{im\Omega t},$$

where $\Omega$ – frequency differences between neighbor comb components.

The density matrix element in the rotation wave approximation has the form:

$$\rho_{12} = \tilde{\rho}_{12}e^{i(\omega_{p0}+kv)t}$$

The density matrix equations are (we omit tilde for slow parts of density matrix elements):

$$\frac{d\rho_{22}}{dt} = \lambda_2 - \gamma_2\rho_{22} + i\frac{d}{2\hbar}\left(\left(W_{1m}e^{i(-2kv)t} + W_{2m}\right)\tilde{\rho}_{21} - \left(W_{1m}^*e^{i2kvt} + W_{2m}^*\right)\tilde{\rho}_{12}\right)$$

$$\frac{d\rho_{11}}{dt} = \lambda_1 - \gamma_1\rho_{11} - i\frac{d}{2\hbar}\left(\left(W_{1m}e^{i(-2kv)t} + W_{2m}\right)\tilde{\rho}_{21} - \left(W_{1m}^*e^{i2kvt} + W_{2m}^*\right)\tilde{\rho}_{12}\right)$$

$$\frac{d\tilde{\rho}_{21}}{dt} + (i(\delta-kv)+\gamma_{21})\tilde{\rho}_{21} = i\frac{d}{2\hbar}\left(W_{1m}^*e^{i2kvt} + W_{2m}^*\right)(\rho_{22}-\rho_{11})$$

We used two rows Fourier expansion method for density matrix elements [12] for the case of moving atoms. Let's write this expansion as it made in [5]:

$$\rho_{ii} = \sum_{l=-\infty}^{\infty}\sum_{p=-\infty}^{\infty} \rho_i^{(l,p)} e^{ilkvt} e^{ip\Omega t}$$

$$\rho_{22} - \rho_{11} = \sum_{l=-\infty}^{\infty}\sum_{p=-\infty}^{\infty} d^{(l,p)} e^{ilkvt} e^{ip\Omega t}$$

$$\tilde{\rho}_{21} = \sum_{l=-\infty}^{\infty}\sum_{p=-\infty}^{\infty} r^{(l,p)} e^{ilkvt} e^{ip\Omega t},$$

$$\tilde{\rho}_{12} = \sum_{l=-\infty}^{\infty}\sum_{p=-\infty}^{\infty} r^{*(-l,-p)} e^{ilkvt} e^{ip\Omega t}.$$

Substituting these expressions in our system of equation, and equating coefficients of like powers, we find

$$d^{(l,\,p)} = \bar{N}\delta_{l,0}\delta_{p,0} + i\frac{d}{2\hbar}D_1^{(l,p)} \times$$

$$\left(\sum_{m=-n}^{n}E_{1m}r^{(l+2,\,p-m)} + \sum_{m=-n}^{n}E_{2m}r^{(l,\,p-m)}\right.$$

$$\left. -\sum_{m=-n}^{n}E_{1m}r^{*(-l+2,-p-m)} - \sum_{m=-n}^{n}E_{2m}r^{*(-l,-p-m)}\right)$$

$$r^{(l,\,p)} = i\frac{d}{2\hbar}D_2^{(l,p)}\left(\sum_{m=-n}^{n}E_{1m}d^{(l-2,\,p+m)} + \sum_{m=-n}^{n}E_{2m}d^{(l,\,p+m)}\right)$$

Here we introduce the function

$$D_1^{(l,p)} = \frac{1}{2}\left(\frac{1}{i(p\Omega + lkv) + \gamma_2} + \frac{1}{i(p\Omega + lkv) + \gamma_1}\right)$$

$$D_2^{(l,p)} = \frac{1}{i(\delta + kv(l-1) + p\Omega) + \gamma_{21}}$$

From these equations, using the condition of reality of diagonal elements $d^{(-l,-p)} = (d^{(l,p)})^*$, we obtain a relation in which as unknowns values remain $d^{(l,\,p)}$ – Fourier components of population difference.

As can be seen from this expression only even members $d^{(l,\,p)}$ are interconnected. To find $d^{(l,\,p)}$ it's necessary to solve an equation of the form **Ad=G**. In this equation, the vector **d** is given by the expression:

$$\mathbf{d} = \begin{pmatrix} d^{(-k_1,\,-k_2)} \\ \vdots \\ d^{(-k_1,\,k_2)} \\ d^{(-k_1+2,\,-k_2)} \\ \vdots \\ d^{(0,\,-1)} \\ d^{(0,\,0)} \\ d^{(0,\,1)} \\ \vdots \\ d^{(k_1-2,\,k_2)} \\ d^{(k_1,\,-k_2)} \\ \vdots \\ d^{(k_1,\,k_2)} \end{pmatrix}$$

A matrix **A** is a matrix of coefficients before to the appropriate components $d^{(l,\,p)}$. Vector G is a column vector $G_i = \bar{N}\delta_{i,((2k_1+1)(2k_2+1)+1)/2}$ with $(2k_1+1)(2k_2+1)$

We write the atomic polarization in the form:

$$P(t) = d(\rho_{21} + c.c.)$$
$$= d\left(\sum_{l=-\infty}^{\infty}\sum_{p=-\infty}^{\infty}r^{(l,\,p)}e^{ilkvt}e^{ip\Omega t}e^{i(\omega_{p0}+kv)t} + c.c.\right)$$

Then for the polarization components at the frequency $\omega_{p0} + kv$ the condition $lkv + p\Omega = 0$ must be satisfied. This condition is satisfied when $l = 0$ and $p = 0$. So, in order to find the polarization at frequency of interest $r^{(0,\,0)}$ must be found.

In an isotropic medium the direction of the polarization vector of the medium coincides with the orientation of the field and for the susceptibility can be written:

$$\chi_{\omega_{p0}} = -\frac{P(\omega)d}{\Omega_{\omega_{p0}}\hbar}$$

Absorption coefficients of the probe field component determined by the imaginary part of the susceptibility:

$$\frac{K_{\omega_{p0}}}{K_\Lambda} = -\frac{2\omega_{21}d}{K_\Lambda}Im(\chi_{\omega_{p0}})$$

where $K_\Lambda = \frac{\omega_{21}d^2}{\hbar\gamma\Gamma}$ - the linear absorption coefficient at the center of the line in the absence of a strong field ($\gamma = \Gamma = 0.5\gamma_{21}$).

We scanned the carrier frequency $\omega_{p0}$ and find the dependence of absorption coefficient from the detuning.

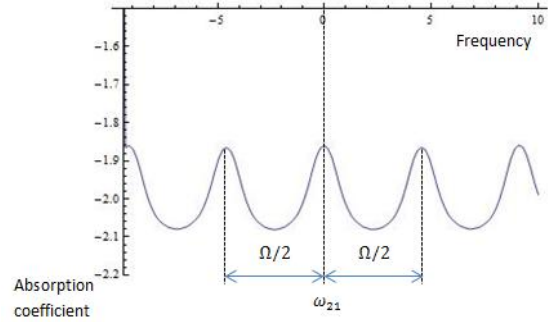

Fig.1. Absorption coefficient in the case of inhomogeneous broadening with rectangular distribution on the velocity ( the width of counter is equal to 10 $\gamma_{12}$
$E_{1m} = E_{2m} = 1$, $\Omega = 10$, $n = 5$.

The shape of inhomogeneous counter for this case is rectangular and the width of counter is equal to 10 $\gamma_{12}$. For Doppler counter shape the Gaussian function is introduced. This case is shown on Fig.2. The width of Doppler counter for this case is equal to 10 $\gamma_{12}$.

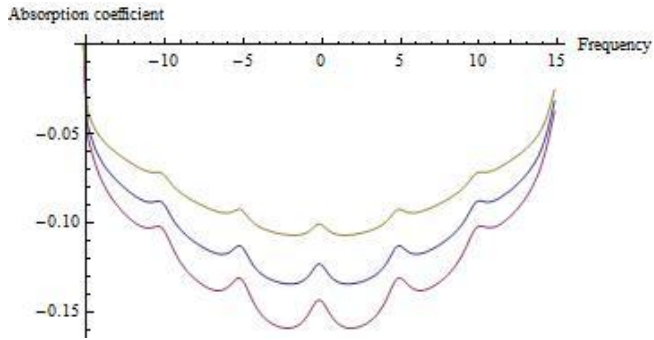

Fig.2. Absorption coefficient in the case of Doppler broadening ( $E_{1m} = E_{2m} = 1; 0.7; 0.5,\ \Omega = 10,\ n = 5$ ).The amplitude of field decreased from lower curve to upper one. The width of Doppler counter for this case is equal to 10 $\gamma_{12}$.

For the strong fields the field broadening of homogeneous line was found. In an absorption spectrum the dips with homogeneous width of the line broadened by a strong field are shown in Fig.2. The contrast of homogeneous dip for strong fields is large than for weak fields.
We check the validity of proposed approach by using direct numerical integration of density matrix equations [11] and found excellent coincidence.

It is shown that the spectral range is defined by a spectral range of radiation of the comb-generator. The spectral range of the comb-generator can be made wide enough for simultaneous registration of a considerable quantity homogeneously - widened spectral lines. It is the big advantage of an offered method comb –spectroscopy on-comparison with existing methods Doppler - free spectroscopy.

All the components of the spectrum of a mixture of gases appear in the spectrum of polarization simultaneously. Doppler shifts which satisfy the conditions of resonance - that is, the Doppler shift must be a multiple of the intermode frequency (difference between adjacent frequency comb - spectrum) or its subharmonics. Thus, the proposed study will have all the advantages of Doppler comb - spectroscopy - high brightness, simultaneous recording in a wide spectral range of a large number of lines of various gases –the new thing is- the resolution within the Doppler contour. These resonances "drawing" the susceptibility spectrum when the carrier frequency of probe comb is scanned along the Doppler shape.

## 3. Conclusions

The susceptibility spectrum of medium of moving atoms driven by counter – propagating waves of strong laser field from generator of chain of femtosecond pulses were researched. It was shown that because multi – photon transitions between quasi – energy levels the waves with frequencies that satisfy of the condition of multi – photon resonance are effective absorbed. The groups of atoms with velocity and wave number that satisfy to the condition of multi-photon resonances are:

$$2kv = (n+l)\Omega.$$

As the result the narrow dip on the background of Doppler shape arises in the absorption spectrum for carrier frequency of probe comb.